\documentclass[11pt,aps,showpacs,preprintnumbers,floatfix]{revtex4}
\usepackage{graphicx}% Include figure files
\usepackage{epsfig}
\usepackage{dcolumn}% Align table columns on decimal point
\usepackage{amsmath}
\def\be{\begin{equation}}
\def\ee{\end{equation}}
\def\bea{\begin{eqnarray}}
\def\eea{\end{eqnarray}}

\def\d#1#2{\frac{\displaystyle #1}{\displaystyle #2}}

\def\p{\partial}
\newcommand{\omits}[1]{}

\def\bsp{\be\begin{split}}

\def\bes{\be  \begin{split}}

\def\p{\partial}

\newcommand{\Rmnum}[1]{\expandafter\@slowromancap\romannumeral #1@}

\def\PRD{{Phys. Rev.}~{\bf D}}
\def\PRL{{Phys. Rev. Lett. }}

\def\PLB{{Phys. Lett.}~{\bf B}}
\def\GRG{{Gen. Rel. Grav. }}
\def\CQG{{Class. Quant. Grav. }}

\def\JHEP{{JHEP}}

\def\IJMPD{{Int. J. Mod. Phys.}~{\bf D}}

\begin{document}

\title{Corrected form of the first law of thermodynamics for regular black holes}
\author{Meng-Sen Ma$^{1,2}$\footnote{Email: mengsenma@gmail.com}, Ren Zhao$^{1,2}$\footnote{Email: zhao2969@sina.com}}

\medskip

\affiliation{\footnotesize$^a$Department of Physics, Shanxi Datong
University,  Datong 037009, China\\
\footnotesize$^b$Institute of Theoretical Physics, Shanxi Datong
University, Datong 037009, China}

\begin{abstract}
We show by explicit computations that there is a superficial inconsistency between the conventional first law of black hole
thermodynamics and Bekenstein-Hawking area law for three types of regular black holes. The corrected form of the first law for these regular black holes is given. The derivation relies on the general structure of the
energy-momentum tensor of the matter fields. When the black hole mass parameter $M$ is included in the energy-momentum tensor, the conventional form of the first law should be modified with an extra factor. In this case, the black hole mass $M$ can no longer be considered as the internal energy of the regular black holes.

\end{abstract}

\pacs{04.70.Dy } \maketitle

\section{Introduction}

The derivation of the laws of black hole mechanics\cite{Hawking:1973}  and the discovery of Hawking radiation\cite{Hawking:1975} suggest a profound connection
between gravitation and thermodynamics. Jacobson's work\cite{Jacobson:1995} by deriving the Einstein equation from a thermodynamic equation of state makes the connection
more closely. The general form of the first law of black hole thermodynamics is
\be\label{1st}
\delta M=T_{H}\delta S+\Omega\delta J +...,
\ee
where the ``..." denote the possible additional contributions from long range fields\cite{Wald:2000}. It has been shown that
this law holds for any field equations derived from a diffeomorphism covariant Lagrangian $L$, which can be written as
\be\label{L}
L=L(g_{ab};R_{abcd},\nabla_{a}R_{bcde}, ...; \psi, \nabla_{a}\psi, ...),
\ee
In the above expression, $\nabla_a$ denotes the covariant derivative associated with $g_{ab}$,
and $\psi$ denotes the collection of all matter fields in the theory.
The entropy corresponding to the Lagrangian, Eq.(\ref{L}), is given by
\be\label{S}
S=-2\pi\oint\d{\p L}{\p R_{abcd}}\epsilon_{ab}\epsilon_{cd}\bar\epsilon.
\ee
The integral is over a slice of the horizon, $\epsilon_{ab}$ is the unit normal bivector to the horizon, and $\bar\epsilon$ is the area element on the horizon slice.
For general relativity,
black hole entropy $S$ takes the form of Bekenstein-Hawking area law $A/4$, namely one quarter
 of the area of black hole event horizon. In general theories of gravity such as Lovelock gravity,
  the black hole entropy is not simply proportional to the area $A$\cite{Myers,TJ,Briscese,Bamba1,Bamba2}.

In this letter, we focus on a special class of black holes, called
regular black holes. As far as we known, there is an inconsistency between the first law and area law for some regular black holes. ``Regular" here means nonexistence of black
hole singularities. More specifically, the metric and the curvature
invariants are all regular everywhere. The first example of a
regular black hole was constructed by Bardeen in 1968\cite{Bardeen}.
Nearly thirty years later, Ay\'{o}n-Beato, et al reobtained the Bardeen
black hole by describing it as the gravitational field of a kind of
nonlinear magnetic monopole\cite{RBH3}. Similarly, many other
regular black holes can also be constructed by introducing nonlinear
electromagnetic sources or phantom field\cite{RBH1,RBH2,EAB:1999,Bronnikov:2001,Bronnikov:2006,Hayward,Dymnikova:2004}.
There is another type of regular black hole, called nonsingular black hole by Dymnikova\cite{Dymnikova:1992,Dymnikova:1996}, with a de Sitter core smoothly connecting to a Schwarzschild
outer geometry. In addition, the noncommutative geometry inspired black holes (NCBH) are also a kind of regular black hole with
the naked singularity replaced by a de Sitter, regular geometry around the origin\cite{Nicolini:2006,Nicolini:2007}.
For details about regular black holes, one can refer to the paper by Lemos et.al\cite{Lemos:2011} and references therein.

Entropy, temperature, Komar energy, Smarr formula, phase transition, etc. of these regular black holes have been studied extensively
\cite{Cheng,Sharif,Myung1,Myung2,Myung3,Rabin:2008,Rabin:2009,Rabin:2011,Nicolini:2011,Smailagic,Dymnikova:2011}.
However, due to the inconsistency between the first law and area law,  these thermodynamic properties are not properly understood.
When studying the thermodynamic properties of these black holes, one has to abandon one law to keep another law to work.
We will give an explanation to the inconsistency and a new understanding of the internal energy of regular black holes.

The paper is organized as follows. In the next section we introduce three types of regular black holes and discuss
 their temperatures, from which one can see the inconsistency between the first law of black hole thermodynamics and the area law mentioned above.
 In Sec.III we give our approach to resolve the inconsistency.  The conclusion is given in Sec.IV.

\par

\section{inconsistency between the first law of black hole thermodynamics and the Bekenstein-Hawking area law}

For a static, spherically symmetric spacetime, the metric in the Schwarzschild gauge can be written as
\be\label{staticmetric}
ds^2=-f(r)dt^2+f(r)^{-1}dr^2 + r^2d\Omega^2
\ee
Below we will only discuss this kind of spacetime for simplicity.

Hawking radiation is just a kinematic effect, which only depends on the event horizon
and is irrelevant to the dynamical equations and the gravitational theories. Therefore, the
temperature of the above black hole can be expressed as
\be\label{tem}
T_h=\d{\kappa}{2\pi}=-\left.\d{1}{4\pi}\d{\p_{r}g_{tt}}{\sqrt{-g_{tt}g_{rr}}}\right|_{r=r_{h}}=
\left.\d{1}{4\pi}f'(r)\right|_{r=r_{h}}
\ee
However, the entropy of the black holes with metric Eq.(\ref{staticmetric}) cannot be determined directly due to its dependence on
the theories of gravity under consideration.

From Eq.(\ref{1st}), it seems that the temperature of a black hole can also be derived according to the entropy
\be\label{MST}
T_H=\left.\d{\p M}{\p S}\right|_{J,...}.
\ee
Or, conversely, the black hole entropy can be calculated according to black hole temperature
\be\label{MTS}
S=\int \left.\d{dM}{T_H}\right|_{J,...}.
\ee
If the first law of black hole thermodynamics is always satisfied, the two temperatures $T_h$ and $T_H$ should be the same.
And the black hole entropy in Eq.(\ref{MTS}) should be the same as that derived for any general covariant using Wald's formula\cite{Wald}.
For most black holes in various gravitational theories it has been well verified that it is indeed the case.
But for some regular black holes, contradiction turns up. Below we will give three examples to demonstrate the problem.

\smallskip

\emph{Bardeen black hole}\cite{RBH3}. For this black hole, the  function $f(r)$ is
\be\label{Bmetric}
 f(r)=1-\d{2Mr^2}{(r^2+g^2)^{3/2}},
 \ee
where the parameters $M,~g$ correspond to mass and magnetic charge of the black hole respectively.
Bardeen black hole is an exact solution of Einstein field equations coupled to a nonlinear magnetic monopole source.
According to Eq.(\ref{S}), it is natural to obtain $S_{bh}=A/4$.
From Eq.(\ref{Bmetric}), one can obtain
\be
M=\frac{\left(g^2+r_h^2\right){}^{3/2}}{2 r_h^2}=\d{\pi(S/\pi+g^2)^{3/2}}{2S},
\ee
where $r_h$ represents the position of event horizon and $S=\pi r_h^2$.
Substituting it into Eq.(\ref{MST}), we can immediately get
\be\label{BardeenMT}
T_H= \frac{\left(S-2 \pi  g^2\right) \sqrt{g^2+S/\pi}}{4 S^2}=\frac{\left(r_h^2-2 g^2\right) \sqrt{g^2+r_h^2}}{4 \pi  r_h^4}.
\ee
While the Hawking temperature derived according to Eq.(\ref{tem}) should be
\be\label{BardeenT}
T_h=\frac{M \left(r_h^3-2 g^2 r_h\right)}{2 \pi  \left(g^2+r_h^2\right){}^{5/2}}=\frac{r_h^2-2 g^2}{4 \pi  g^2 r_h+4 \pi  r_h^3}.
\ee
Obviously, the two temperatures are different generally, which means the inconsistency between the area law and the first law of black hole thermodynamics.
When $g=0$, they will  coincide and agree with the temperature of Schwarzschild black hole.

Some authors choose to start with the Hawking temperature, Eq.(\ref{BardeenT}), and employ Eq.(\ref{MTS}) to derive the entropy\cite{Cheng,Sharif}.
The result is
\be
S=\d{\pi}{r_h}(r_h^2-2g^2)\sqrt{r_h^2+g^2}+3\pi g^2\ln(r_h+\sqrt{r_h^2+g^2}).
\ee
Although this entropy fulfills the firs law, it lacks a reasonable explanation.
In fact, the result does not only violate Wald's formula, Eq.(\ref{S}), but also Visser's result\cite{Visser}.

\smallskip
\emph{Noncommutative geometry inspired Schwarzschild black hole}\cite{Nicolini:2006}(NCSBH). The $f(r)$ takes the form
\be
f(r)=1-\d{4M\gamma(3/2,r^2/4\theta)}{r\sqrt{\pi}},
\ee
where $\theta$ is a constant with dimension of length squared and represents the noncommutativity of spacetime.
$\gamma(3/2,r^2/4\theta)$ is the lower incomplete gamma function with definition
  \be
  \gamma(3/2,r^2/4\theta)=\int_0^{r^2/4\theta}dt~ t^{1/2}e^{-t}.
  \ee
 Noncommutativity is expected to be relevant
at the Planck scale where the usual semiclassical considerations break down. Thus, one point of view is that the Bekenstein-Hawking area law does not apply to the NCBH.
One should start from the first law of black hole thermodynamics and derive the black hole entropy according to Eq.(\ref{MTS}).
In \cite{Myung3,Rabin:2008,Rabin:2009,Rabin:2011,Nicolini:2011,Smailagic}, this work is done and indeed an entropy different from $A/4$ is given.

It has been remarked in \cite{Nicolini:2006} that the noncommutative effects can be implemented acting only on
the matter source, and the Einstein tensor part of the field equations will leave unchanged. Specifically, the microscopic degree of freedom of
spacetime( such as noncommutativity) will play a role only when spacetime is probed at Planck scale. Therefore, we  may consider the entropy of the NCSBH black hole
to accord with the Bekenstein-Hawking area law.

One can express the black hole mass as
\be
M=\d{\sqrt{\pi}r_{h}}{4\gamma(3/2,r_{h}^2/4\theta)}=\d{\sqrt{S}}{4\gamma(3/2,S/4\pi\theta)},
\ee
from which we can employ the first law of black hole thermodynamics to calculate
\be
T_H=\d{\p M}{\p S}=\d{A(r_{h},\theta)}{8\sqrt{\pi}r_{h}\gamma(3/2,r_{h}^2/4\theta)}
%&=&\d{1}{8\sqrt{S}\gamma(3/2,S/4\pi\theta)}\left[1-\d{1}{4}\sqrt{\left(\d{S}{\pi\theta}\right)^3}\d{e^{-S/4\pi\theta}}{\gamma(3/2,S/4\pi\theta)}\right]
\ee
where $A(r_{h},\theta)=1-\d{r_{h}^3}{4\theta^{3/2}}\d{e^{-r_{h}^2/4\theta}}{\gamma(3/2,r_{h}^2/4\theta)}$.
\par
On the other hand
\be
T_h=\left.\d{1}{4\pi}f'(r)\right|_{r=r_{h}}=\d{A(r_{h},\theta)}{4\pi r_{h}}.
\ee
There is also a discrepancy between the two temperatures.

\par
\emph{Dymnikova's nonsingular black hole}\cite{Dymnikova:1992}.
\be
f(r)=1-\d{2M}{r}\left(1-e^{-\d{r^3}{2r_0^2M}}\right),
\ee
where the parameter $r_0$ is a constant and is connected with $\epsilon_0$, the vacuum energy density, by the de Sitter relation $r_0^2=3/8\pi\epsilon_0$.

For the nonsingular black hole,
\be\label{nonh}
T_h=\d{3 r_h^3 -3 r_h^2r_g+r_0^2 r_g}{4\pi r_{h}r_{g}r_{0}^2},
\ee
where $r_{g}=2M$.
According to \cite{Dymnikova:1992,Dymnikova:1996,Dymnikova:2011}, there is no doubt that the entropy of the nonsingular black hole takes the form of Bekenstein-Hawking area law.
From the metric, one should solve the equation
\be
r_{h}=2M\left[1-\exp\left({\d{-r_{h}^3}{2Mr_{0}^2}}\right)\right]
\ee
to obtain
\bea
M=\frac{r_h^3}{2 r_0^2 W\left(-\frac{ r_h^2}{r_0^2}~e^{-r_h^2/r_0^2}\right)+2 r_h^2}= \frac{S^{3/2}}{ 2 \pi^{3/2}  r_0^2 W\left(-\frac{S}{\pi  r_0^2}~e^{-S/\pi  r_0^2}\right)+2\sqrt{\pi } S}
\eea
where $W(x)$ is the Lambert function defined by the general formula $W(x)e^{W(x)}=x$. Therefore,
\be\label{nonms}
T_H=\d{\p M}{\p S}=\frac{r_g \left(3 r_h^3 -3 r_h^2r_g+r_0^2 r_g\right)}{4 \pi  r_h^2 \left(r_h^3 -r_h^2r_g+r_0^2 r_g\right)}
\ee
Once again, the two temperatures are different.

One can see from the above examples that there are two temperatures or two entropies. If choosing to believe in the Bekenstein-Hawking area law,
we have to abandon  the first law of black hole thermodynamics due to the wrong temperature derive from Eq.(\ref{MST}). On the other hand, if we require
that the first law is satisfied with the regular black holes, the entropy is no longer the Bekenstein-Hawking one.

Besides the above regular black holes, some others \cite{RBH1,RBH2,EAB:1999,Bronnikov:2001,Bronnikov:2006,Dymnikova:2004} also violate the first law of black hole thermodynamics in this way.
The regular black holes in the above examples are all solutions of Einstein field equations. Thus, we tend to think that the black hole entropy should take the form of
 Bekenstein-Hawking area law. So, the question we would like to address is how to reconcile the contradiction
 between the first law and the area law for those black holes mentioned above.

In addition, there is another interesting point. According to the work\cite{Jacobson:1995}, one can consider another form of the first law, $\delta Q=T\delta S$.
If defining
\be
\delta Q=\int_{H}T_{ab}\chi^{a}d\Sigma^{b}
\ee
and replacing $T_{ab}$ with the gravitational field equation as what is done in the paper\cite{Li}, the relation $\delta Q=T\delta S$,
even for the regular black holes mentioned above, is satisfied obviously.

\par

\section{internal energy for the regular black holes}

In fact, the black holes mentioned above are not \textbf{vacuum} solutions of Einstein field equations.
Although the nonsingular black hole\cite{Dymnikova:1992} is titled `` vacuum nonsingular black hole " by the author,
it is derived from the Einstein field equations with nonzero energy-momentum tensor. In \cite{Ortin}, the matter source problem of gravitational field has been discussed elaborately. Although a completely understanding on the problem, specially at $r=0$, is still lacking.  One can find that the key to solve the contradiction between the first law and the area law lies in taking into account the contributions of the matter fields.

According to the discussion in \cite{Hawking:1973}, for asymptotically flat spacetimes the integral and the differential mass formulae are respectively
\be\label{1sti}
M=\int_{S}(2T_{\nu}^{~\mu}-T\delta_{\nu}^{~\mu})\xi^{\nu}_{~(t)}d\Sigma_{\mu}+2\Omega J+\d{\kappa}{4\pi}A,
\ee
and
\be\label{o1st}
\delta M=\delta\int_{S} T_{\nu}^{~\mu}\xi^{\nu}_{~(t)}d\Sigma_{\mu}+ \Omega\delta J+\d{\kappa}{8\pi}\delta A.
\ee
When this is applied to the case of linear or nonlinear electrodynamics, even with scalar fields, the first law also holds true and the first term on the
right hand side will be replaced with $\Phi\delta Q+...$\cite{Rasheed,Gibbons}. But, if the energy-momentum tensor $T_{\mu\nu}$ includes
the black hole mass $M$, the situation will be very different.

\omits{
\be\label{deltaM}
\delta M=4\pi\delta\int_{S} r^2T_{~0}^{0}dr+\d{\kappa}{8\pi}\delta A
\ee

For a static, spherically symmetric spacetime, the two mass formulae above become
\be\label{s1}
M=4\pi\int_{r_h}^{\infty}(T-2T_{~0}^{0})r^2dr+\d{\kappa}{4\pi}A
\ee
and
\be\label{s2}
\delta M=-4\pi\delta\int_{r_h}^{\infty} r^2T_{~0}^{0}dr+2\pi\int_{r_h}^{\infty}T^{\alpha\beta}\delta g_{\alpha\beta}r^2dr +\d{\kappa}{8\pi}\delta A
\ee

}

Below we will derive the differential mass formula from the Einstein field equations.
We assume the function $f(r)$ in Eq.(\ref{staticmetric}) to take the form of
\be\label{efm}
f(r)=1-\d{2m(r)}{r},
\ee
where $m(r)$ is an effective mass. It is clear that the event horizon lies at the position where $m(r_{h})=r_{h}/2$.
According to Einstein field equations, $m(r)$ should satisfy
\be\label{mr}
\d{d m(r)}{d r}=-4\pi r^2 T^0_{~0}.
\ee
Integrating the above equation, one can obtain
\be\label{mr2}
m(r)=M+4\pi\int_{r}^{\infty}r^2T^0_{~0}dr,
\ee
where the integration constant has been taken to be the black hole mass $M$ because $M=\lim\limits_{r\rightarrow \infty}m(r)$.
Substituting Eqs.(\ref{efm}),(\ref{mr2}) into Eq.(\ref{tem}), we can get the Hawking temperature of the black holes
\be
T_h=\d{1}{4\pi r_h}+2 r_h T_{~0}^{0}
\ee
At the horizon, one can derive from Eq.(\ref{mr2})
\be
M=\d{r_{h}}{2}-4\pi\int_{r_{h}}^{\infty}r^2T^0_{~0}dr
\ee
Taking the variation of the above equation, one can obtain
\be
\delta M=\d{1}{2}\delta r_{h}-4\pi \delta \int_{r_{h}}^{\infty}r^2T^0_{~0}dr.
\ee
It should be noted that if $T^0_{~0}=T^0_{~0}(r,M)$, namely $T^0_{~0}$ is not only a function of $r$ but also the black hole mass $M$, one can obtain
\be
\left(1+4\pi\int_{r_{h}}^{\infty}r^2\d{\p T^0_{~0}}{\p M}dr\right)\delta M=\left(\d{1}{4\pi r_h}+2r_{h}T_{~0}^{0}\right)\delta \d{A}{4}\nonumber,
 \ee
 namely
 \be\label{result}
 C(M,r_h)\delta M=T_h~\delta \d{A}{4}.
 \ee
In fact, this should be the first law of black hole mechanics.
Due to the above equation, if the Bekenstein-Hawking area law, namely $S=A/4$, is satisfied, the conventional first law of thermodynamics must be
violated for the regular black holes we studied.
Obviously,  the form of the first law of black hole thermodynamics is different from the conventional one with an extra factor.
Therefore, if the entropy of the regular black holes satisfy the area law, the temperature should be
\bea\label{CT}
T_h=\left(1+4\pi\int_{r_{h}}^{\infty}r^2\d{\p T^0_{~0}}{\p M}dr\right)\d{\p M}{\p S}=C(M,r_h)\d{\p M}{\p S}=C(M,r_h)T_{H}
\eea
It is clear that for the conventional cases with $T^0_{~0}$ independent of $M$, the  factor $C(M,r_h)=1$, the Eq.(\ref{result}) coincides with the first law of black hole thermodynamics without sources. The two temperatures $T_h$ and $T_H$ are also the same.

Several comments are needed for the result, Eq.(\ref{result}). Because any non-extremal black hole has temperature, it can be seemed as a thermodynamic system.
Thus, the conventional thermodynamic laws must be satisfied. We have two choices to connect  Eq.(\ref{result}) with the first law of thermodynamics, $\delta E=T\delta S$. The first one is $E\leftrightarrow M, ~\delta S \leftrightarrow \delta \d{A}{4}/C(M,r_h), ~T\leftrightarrow T_h$; The second one is $S\leftrightarrow A/4, ~\delta E \leftrightarrow C(M,r_h)\delta M,~T\leftrightarrow T_h $. We tend to believe the latter one. The reason is due to Wald's formula, from which the entropy of the regular black holes we studied should satisfy the area law. Thus the black hole mass $M$ cannot be considered as the internal energy $E$ of the black hole system generally.
Eq.(\ref{result}) can be easily generalized to charged black holes. In the case with $T_{~0}^{0}=T_{~0}^{0}(r,M,Q)$, the first law will be
expressed as
\be
C(M,Q,r_h)\delta M=T_h\delta S+\Phi_h\delta Q,
\ee
where $C(M,Q,r_h)=\left(1+4\pi\int_{r_{h}}^{\infty}r^2 \left.\d{\p T^0_{~0}}{\p M}\right|_Q dr\right)$
and the electric potential on the horizon can be calculated according to electric field, $\Phi_h=\int_{r_h}^{\infty}E dr$.
Some more terms can appear on the right hand side due to the presence of other matter fields.
Since our derivation depends on the static, spherically symmetric spacetime, it would be interesting to repeat our calculation for rotating black holes.

Padmanabhan had ever found that  for
a spherically symmetric spacetime the Einstein's equations can be interpreted as the thermodynamic identity\cite{Pa1,Pa2,Pa3}
\be\label{eos}
dE=TdS-PdV,
\ee
where $E=\left(\d{A_H}{16\pi}\right)^{1/2}, T=T_h,\quad S=\d{A_H}{4},\quad P=-T_r^{~r}(r_h),\quad V=\d{4}{3}\pi r_h^3$. The contributions from matter fields have been contained in the pressure $P$. Obviously, except for vacuum cases,
$E\neq M$ generally. In this way, one can also find that for General relativity the black hole entropy should satisfy the Bekenstein-Hawking area law, at least in static sperically symmetric spacetime. The internal energy proposed by Padmanabhan for the black hole thermodynamic system is different from our result. Padmanabhan considers black hole thermodynamic system as a closed system which includes all the matter fields near the black hole. While we still follow Bekenstein's viewpoint and consider the contributions from the matter fields around the black hole as chemical potentials.

Now we can reconsider the several types of regular black holes mentioned before. The Bardeen black hole can be a magnetic solution to
Einstein equations coupled to a nonlinear electrodynamic source, which has the Lagrangian
\be
\textit{L}(F)=\d{3}{2sg^2}\left(\d{\sqrt{2g^2F}}{1+\sqrt{2g^2F}}\right)^{5/2}=\d{3Mg^2}{(r^2+g^2)^{5/2}},
\ee
where $s\equiv |g|/2M$, and $F\equiv \d{1}{4}F_{\mu\nu}F^{\mu\nu}=g^2/2r^4$. One can easily find out $T_{~0}^{0}=-L(F)/4\pi$. Thus,
 \be
 C(M,g,r_h)=1-3g^2\int_{r_{h}}^{\infty}\d{r^2}{(r^2+g^2)^{5/2}}dr=\d{m(r_h)}{M}
 \ee
Substituting the above result and Eq.(\ref{BardeenMT}) into Eq.(\ref{CT}), one can easily derive Eq.(\ref{BardeenT}).

For the NCSBH,
\be
T_{~0}^{0}=-\rho_{\theta}=-\d{M}{(4\pi\theta)^{3/2}}\exp\left(-r^2/4\theta\right)
\ee
one can compute the  factor
\bea
C(M,r_h)=\d{2}{\sqrt{\pi}}\gamma(3/2,r_{h}^2/4\theta)=\d{m(r_h)}{M}
 \eea
Substituting this result into Eq. (\ref{CT}), one can also find that the Hawking temperature restores. Therefore, the semiclassical entropy of the NCBH
still takes the form of area law.
From the above two examples, it seems that
the corrected factor is just $m(r_{h})/M$. In fact, we cannot draw this conclusion rashly. The $T_{~0}^{0}$ of the Bardeen black hole and the NCSBH are both proportional to the black hole mass $M$.
For nonsingular black hole, the $``00"$ component of energy-momentum tensor is given by
\be
T_{~0}^{0}=-\d{3}{8\pi r_{0}^2}\exp\left(\d{-r^3}{2Mr_{0}^2}\right)
\ee
Thus
\bea
 C(M,r_h)=1-\d{r_{0}^2r_{g}+r_{h}^3}{r_{g}r_{0}^2}\exp\left(\d{-r^3}{2Mr_{0}^2}\right)=\d{r_{h}^4+r_{h}r_{g}r_{0}^2-r_{h}^3r_{g}}{r_{0}^2r_{g}^2}
\eea
It can be easily verified that the result in Eq.(\ref{nonms}) multiplied by the factor can give Eq.(\ref{nonh}).

\omits{
According to Eq.(\ref{mr2}), one can demonstrate that $T_{~0}^{0}$ as a function of black hole mass $M$ is a prerequisite condition for the regular black holes.
For the regular black holes, there must be $\lim\limits_{r\rightarrow 0}m(r)=0$. Therefore,
\be
4\pi\int_{0}^{\infty}r^2T_{~0}^{0}dr =M
\ee
Thus, $T_{~0}^{0}$ must be a function of $M$, or else one will obtain $M$ as a constant.}

\par

\section{Concluding remarks}

To summarize, the inconsistency between the first law of black hole thermodynamics and the Bekenstein-Hawking area law for some regular black holes
originates from the inclusion of the black hole mass $M$ in the energy-momentum tensor, which induces an extra factor, $C(M,...)$, in the conventional used first law. In the semiclassical approximation, this conclusion does not only apply to the regular black holes mentioned above, but also all the other black holes.
In fact, this is a more general and complete form. It will return to the conventional form of the first law when the energy-momentum tensor is independent of the black hole mass explicitly. Specifically, for the regular black holes, which are exact solutions of Einstein field equations, the black hole mass $M$ in general cannot be equated to the internal energy of the thermodynamic system of black hole. With the redefinition of the internal energy of regular black hole,
the first law of thermodynamics and the Bekenstein-Hawking area law are consistent.

\bigskip

\section*{Acknowledgements}
The author would like to thank Professor Chao-Guang Huang for useful discussion.
This work is supported in part by the National Natural Science Foundation of China under Grants
No.11247261, No.11075098 and by the Doctoral Sustentation
Fund of Shanxi Datong University (2011-B-03).

\end{document}